\documentclass[showpacs,preprintnumbers,amsmath,amssymb]{revtex4}
\usepackage{amscd}
\usepackage{amsthm}
\usepackage{color}
\usepackage{graphicx}
\usepackage{hyperref}
\usepackage{floatrow}
\usepackage{cases}

\def\qed{\leavevmode\unskip\penalty9999 \hbox{}\nobreak\hfill
     \quad\hbox{\leavevmode  \hbox to.77778em{%
               \hfil\vrule   \vbox to.675em%
               {\hrule width.6em\vfil\hrule}\vrule\hfil}}
     \par\vskip3pt}
\begin{document}

\title{\Large {\bf LOCC distinguishable orthogonal product states with least entanglement resource}}

\author{Haiquan Li$^{1}$, Xilin Tang$^{1\ast}$, Naihuan Jing$^{1, 2}$, Ze Gu$^3$}

 \affiliation
 {
 {\footnotesize  {$^1$Department of Mathematics,
 South China University of Technology, Guangzhou
510640, P.R.China}} \\
{\footnotesize{
  $^2$Department of Mathematics, North Carolina State University,
Raleigh, NC 27695, USA}}\\
{\footnotesize{$^3$School of Mathematics and Statistics, Zhaoqing University, Zhaoqing, Guangdong, China}}
}

\begin{abstract} \label{abstract}
In this paper, we construct $2n-1$ locally indistinguishable orthogonal product states in
$\mathbb{C}^n\otimes\mathbb{C}^{4}~(n>4)$ and $\mathbb{C}^n\otimes\mathbb{C}^{5}~(n\geq 5)$ respectively. Moreover, a set of locally indistinguishable orthogonal product states with $2(n+2l)-8$ elements in $\mathbb{C}^n\otimes\mathbb{C}^{2l}~(n\geq 2l>4)$ and a class of locally indistinguishable orthogonal product states with $2(n+2k+1)-7$ elements in $\mathbb{C}^n\otimes\mathbb{C}^{2k+1}~(n\geq 2k+1>5)$ are also constructed respectively.
These classes of quantum states are then shown to be distinguishable by local operation and classical communication (LOCC) using a suitable $\mathbb{C}^2\otimes\mathbb{C}^2$ maximally entangled state respectively.
\end{abstract}

\maketitle
\maketitle

\section{Introduction}

\bigskip

Quantum entanglement is both mysterious and useful in quantum information processing due to its wide range of applications
in quantum cryptography\cite{Ekert91, Gisin02}, quantum teleportation\cite{Karlsson98, Kim01}, and quantum secured direct communication \cite{Wang05, Tian08, Yang11} and so on. Its mystery has spurred investigations on other closely related quantum phenomena such as
quantum nonlocality, as local operation and classical communication (LOCC) protocol also plays an important role in quantum information theory.

LOCC is often used as a tool to verify whether quantum states are perfectly distinguishable or not \cite{Fan04, Nathanson05, Nathanson13}.  Nonlocality of quantum information is revealed when a set of orthogonal states cannot be distinguished by LOCC. This problem underpins quantum entanglement as a resource.
 During the past twenty years, quantum nonlocality has been studied intensively.
 Given a known set of mutually orthogonal states, two separated observers Alice and Bob share a bipartite quantum system but they don't know which state their combined system is in. One says that these states are locally distinguishable if Alice and Bob can reliably determine them
 by LOCC. There are many interesting results on LOCC distinguishability \cite{Ghosh01, Cohen007, Duan07, Band09, Yu11, Zhang015} of orthogonal states and LOCC indistinguishability of orthogonal states \cite{Horodecki03, Feng09, Zhang013}.

It is well known that entanglement is not necessary for local indistinguishable quantum states. Bennett et al.\cite{Ben99} first presented nine LOCC indistinguishable orthogonal product states in $\mathbb{C}^3\otimes\mathbb{C}^3$ and showed the phenomenon of nonlocality without entanglement.
Walgate and Hardy\cite{Walgate02} gave a simpler method, in which a necessary and sufficient condition for distinguishability of orthogonal states in $\mathbb{C}^2\otimes\mathbb{C}^n$ was used. Zhang et al.\cite{Zhang14} extended the result and constructed $d^{2}$ local indistinguishable orthogonal product states in $\mathbb{C}^d\otimes\mathbb{C}^d$, where $d$ is odd.
They further proved that there exist $4d-4$ orthogonal product states that cannot be perfectly distinguished in $\mathbb{C}^d\otimes\mathbb{C}^d$\cite{zhang15}. Recently Wang et al.\cite{Wang15} showed that there are 
$3(m+n)-9$ orthogonal product states in $\mathbb{C}^m\otimes\mathbb{C}^n(3\leq m\leq n)$ and they are LOCC indistinguishable, and the authors further
obtained a set of $3n+m-4$ orthogonal product states in $\mathbb{C}^m\otimes\mathbb{C}^n(3\leq m\leq n)$ \cite{Zhang16} and proved that they are perfectly distinguished by LOCC. Most recently Xu et al.\cite{Xu15} constructed $2p-1$ locally indistinguishable orthogonal product states
in $\mathbb{C}^m\otimes\mathbb{C}^n$ ($3\leq m\leq n, 3\leq p\leq m)$. Despite these interesting developments, locally indistinguishable orthogonal product states are unknown for the most general bipartite system.

On the other hand, locally indistinguishable quantum states may become distinguishable by LOCC via entanglement.
For a set of orthogonal states in some bipartite system undistinguished by LOCC, it is possible to distinguish them when Alice and Bob share an entangled state. In $2008$, Cohen presented an effective method to perfectly distinguish certain classes of unextendible product bases (UPB) in $\mathbb{C}^m\otimes\mathbb{C}^n$ using entanglement \cite{Cohen07}. In 2016, Zhang et al. used the method to distinguish indistinguishable orthogonal product states in $\mathbb{C}^m\otimes\mathbb{C}^n$ with a $\mathbb{C}^2\otimes\mathbb{C}^2$ maximally entangled state \cite{Zhang016}. However, not all orthogonal product states are LOCC distinguishable by using only a $\mathbb{C}^2\otimes\mathbb{C}^2$ maximally entangled state.
It is necessary to consider other entanglement resources and obtain
optimal ones to distinguish a given set of orthogonal product states.

In this paper, we will construct a set of locally indistinguishable orthogonal product states in $\mathbb{C}^m\otimes\mathbb{C}^n$ and
show that they are local distinguishable using entanglement as a resource. 

The paper is organized as follows. In Sec. 2, we first construct $2n-1$  orthogonal product states in $\mathbb{C}^n\otimes\mathbb{C}^{4}~(n>4)$ and $2(n+2l)-8$ orthogonal product states in $\mathbb{C}^n\otimes\mathbb{C}^{2l}~(n\geq 2l>4)$ and show that they are local indistinguishable respectively. Then we consider $\mathbb{C}^n\otimes\mathbb{C}^{2k+1}(n\geq 2k+1\geq 5 )$. When $n\geq 2k+1=5$, we find $2n-1$ locally indistinguishable orthogonal product states. When $n\geq 2k+1>5$, there are $2(n+2k+1)-7$ orthogonal product states that cannot be perfectly distinguished by LOCC. In Sec. 3, we prove that these indistinguishable states can be distinguished by LOCC with a $\mathbb{C}^2\otimes\mathbb{C}^2$ maximally entangled state respectively, which is obviously the least entanglement resource for the proof of method of Cohen in article\cite{Cohen07}.
Discussions and summary are given in Sec. 4.

\bigskip
\section{ Constructions of LOCC indistinguishable orthogonal product states }

In this section, we construct locally indistinguishable orthogonal product states in the system $\mathbb{C}^n\otimes\mathbb{C}^m(n \geq m \geq 4)$ according to $m$ being even or odd. First we fix some terms to simplify presentation.

Let $\{|i\rangle\}_{i=1}^{d}$ be a fixed orthonormal basis in $\mathbb{C}^d$, and we also consider the states $|i\pm j\rangle=\frac{1}{\sqrt{2}}(|i\rangle\pm|j\rangle)$. If there are two parties called Alice and Bob, and
Alice performs a nontrivial measurement upon the system first, then we say that Alice goes first \cite{Walgate02} .

We call a measurement trivial if all the POVM elements are proportional to the identity operator, as such a measurement yields no information about the state. All other measurements will be regarded as nontrivial.

A pure state $|\psi\rangle$ is called a $\mathbb{C}^{d}\otimes\mathbb{C}^{d'} (d' > d)$ maximally entangled state (MES) if for arbitrary given orthogonal complete basis ${|i_{A}\rangle}$ of the subsystem $A$, there exists an orthogonal basis ${|i_{B}\rangle}$ of the subsystem $B$ such that $|\psi\rangle$ can be written as $|\psi\rangle=\frac{1}{\sqrt{d}}\sum\limits_{i=1}^{d}|i_{A}i_{B}\rangle$ \cite{Li12}.

To present our results better, we use box-diagrams to show mutually orthogonal product states
as in \cite{Cohen07} given by Cohen. Three examples
of LOCC indistinguishable orthogonal product states are depicted in Fig.1, Fig.2 and Fig.3, in which
a tile represents a product state.
As an example, an arbitrary state $\varepsilon$ on system A and system B is in the state $|i\rangle$ and in the states $|j\rangle$ and $|j+1\rangle$, respectively. We can write $\varepsilon$ as $\varepsilon=|i\rangle_A\otimes|j+(j+1)\rangle_B$.
An exception is that the two squares labeled $7$ in Fig.1 represent the
state $|3-5\rangle_A|2\rangle_B$ in system $\mathbb{C}^6\otimes\mathbb{C}^4$. In addition, the stopper state (i.e. $|\phi\rangle$) is not shown in the figures, as it would cover the whole diagram.

\begin{figure}[ht]
 \floatsetup{floatrowsep=qquad}
 \begin{floatrow}
 \ffigbox[\FBwidth]{\caption{Tiling structure of orthogonal product states in $\mathbb{C}^6\otimes\mathbb{C}^4$. }\label{bipartite_Example1}}{\includegraphics[width=0.46\textwidth,height=0.4\textwidth]{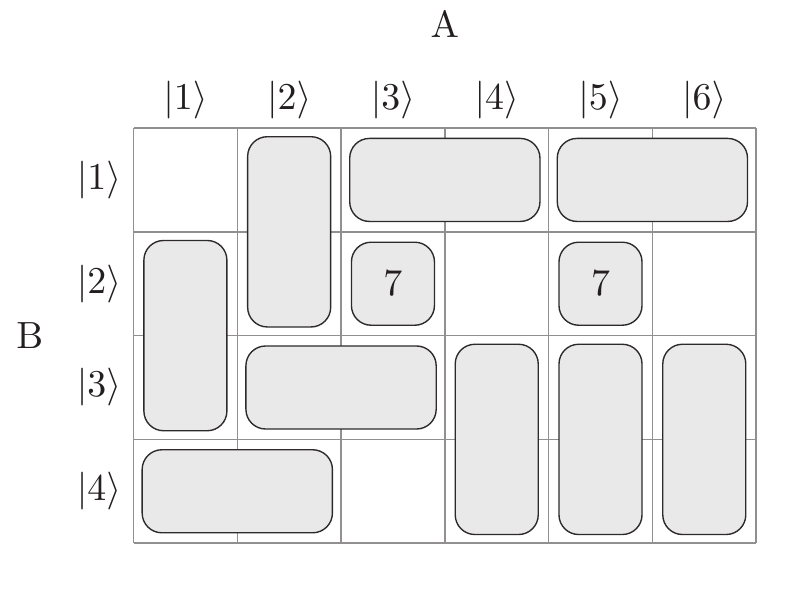}}

\ffigbox[\FBwidth]{\caption{Tiling structure of orthogonal product states in $\mathbb{C}^7\otimes\mathbb{C}^6$. }\label{bipartite_Example1}}
{\includegraphics[width=0.52\textwidth,height=0.5\textwidth]{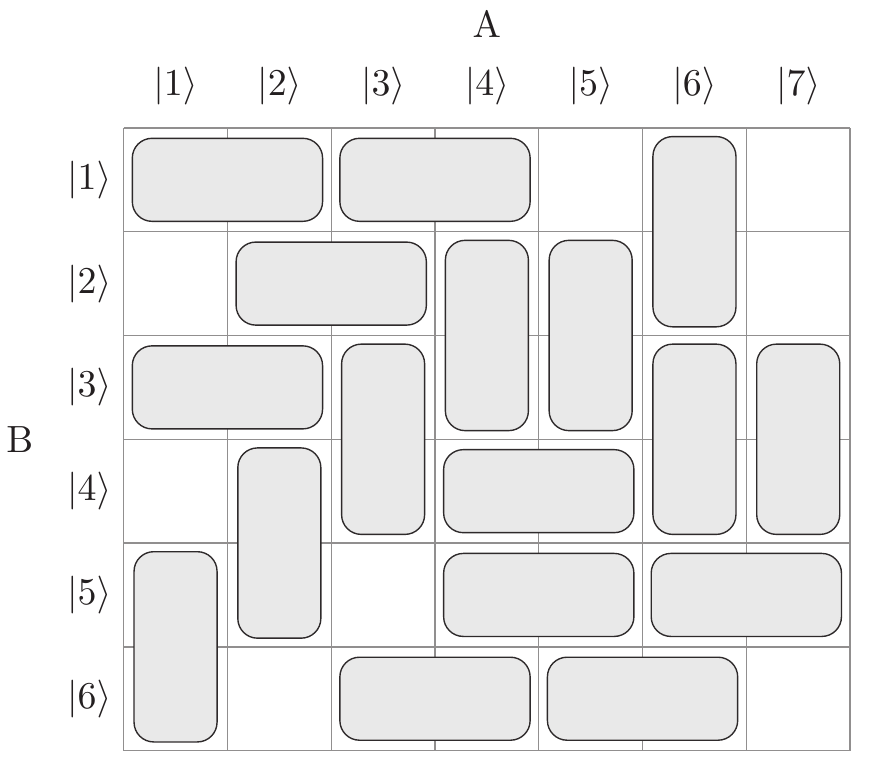}
}
 \end{floatrow}
\end{figure}

\begin{figure}
  \centering
  \caption{Tiling structure of orthogonal product states in $\mathbb{C}^9\otimes\mathbb{C}^7$.}
  \includegraphics{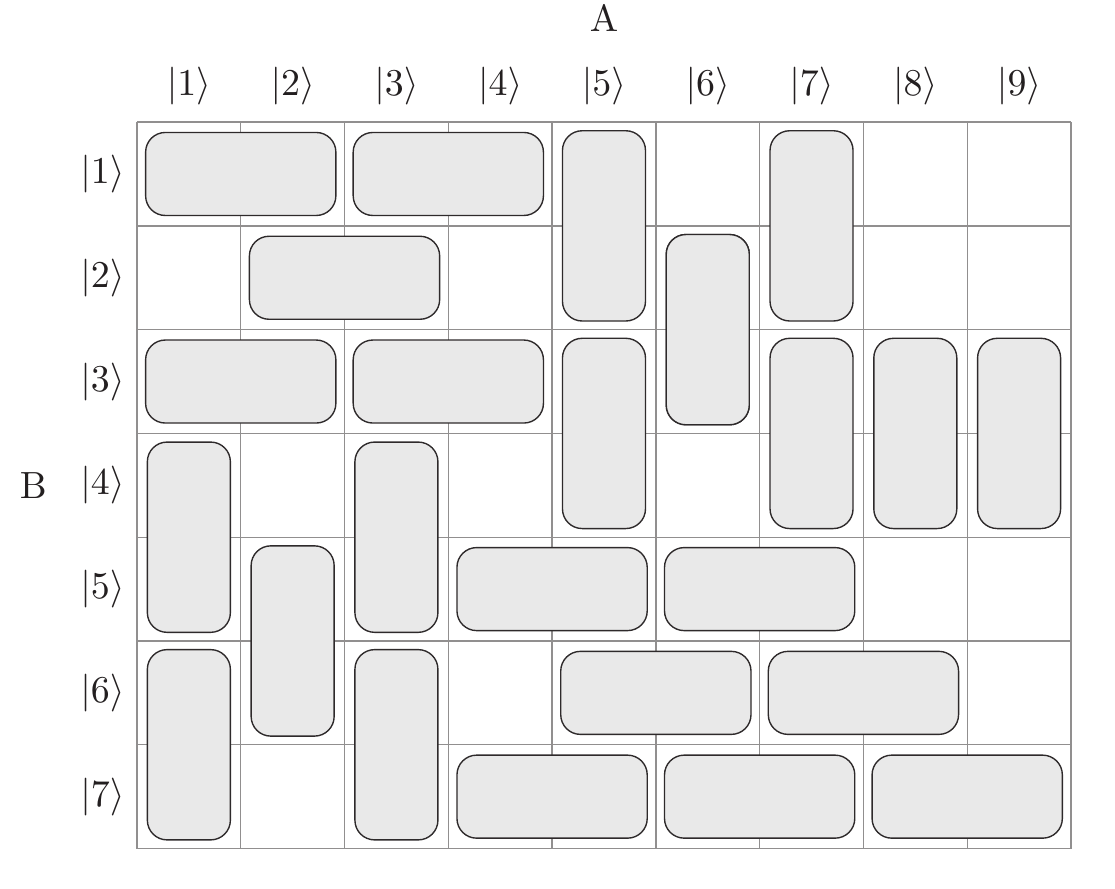}
\end{figure}

Case 1. In $\mathbb{C}^n\otimes\mathbb{C}^m$ ($m=2l$), we find $2(m+n)-9$ LOCC indistinguishable orthogonal product states in case $n>m=4$ and $2(m+n)-8$ LOCC indistinguishable orthogonal product states in case $n\geq m>4$, respectively.

\bigskip
\noindent {\bf Theorem 1.}  In $\mathbb{C}^n\otimes\mathbb{C}^m$ $(n > m = 4),$ there are $2n-1$ orthogonal product states that are LOCC indistinguishable and
 are constructed in (1).

\begin{equation}
\begin{split}
&|\phi\rangle=|1-2+\cdots+(-1)^{n-1}n\rangle_A|1-2+3-4\rangle_B,\\
&|\varphi_1\rangle=|1\rangle_A|2+3\rangle_B,~~~|\varphi_2\rangle=|2+3\rangle_A|3\rangle_B, \\
&|\varphi_3\rangle=|1+2\rangle_A|4\rangle_B,~~~|\varphi_4\rangle=|2\rangle_A|1+2\rangle_B, \\
&|\varphi_5\rangle=|3+4\rangle_A|1\rangle_B, \\
&|\varphi_{i+5}\rangle=\begin{cases} |(4+i)+(5+i)\rangle_A|1\rangle_B,i=1,3,\ldots, 2\lfloor\frac{n-4}{2}\rfloor-1, \\
         |(4+i)+(5+i)\rangle_A|2\rangle_B,i=2,4,\ldots, 2\lceil\frac{n-4}{2}\rceil-2 \end{cases},\\
&|\varphi_{n+1}\rangle=|3-5\rangle_A|2\rangle_B,\\
&|\varphi_{i+n+2}\rangle=|(4+i)\rangle_A|3+4\rangle_B,i=0,1,2,\ldots, n-4. \\
\end{split}
\end{equation}

\noindent{ Proof:}
We prove that the orthogonal product states in (1) are indistinguishable by LOCC no matter who goes first.
We start by assuming that Alice goes first with a non-disturbing measurement $\{M^{A}_{m}\}$. It is enough to show that for any matrix $M^{A}_{m}$ preserving the orthogonality of given states, i.e. $M^{A}_{m}\otimes I|\xi_{i}\rangle$ are mutually orthogonal when $|\xi_{i}\rangle$ belong to (1), $M^{A^{\dag}}_{m}M^{A}_{m}$ is proportional to the identity. Write $M^{A^{\dag}}_{m}M^{A}_{m}=(a_{ij})_{n\times n}$.  Since
$\{ M^{A}_{m}\otimes I|\xi_{i}\rangle \}$ is an orthogonal set, $a_{ij}=0$ for $i\neq j$. This can be seen by an inductive argument:

\begin{equation}\label{e:1}
\begin{split}
&0=\langle \varphi_1|M^{A^{\dag}}_{m}M^{A}_{m}\otimes I|\varphi_4\rangle=\langle 1|M^{A^{\dag}}_{m}M^{A}_{m}|2\rangle=a_{12},\\
&0=\langle \varphi_1|M^{A^{\dag}}_{m}M^{A}_{m}\otimes I|\varphi_2\rangle=\langle 1|M^{A^{\dag}}_{m}M^{A}_{m}|2+3\rangle=a_{12}+a_{13}=a_{13},\\
&0=\langle \varphi_1|M^{A^{\dag}}_{m}M^{A}_{m}\otimes I|\varphi_{i+n+2}\rangle=\langle 1|M^{A^{\dag}}_{m}M^{A}_{m}|(4+i)\rangle=a_{1(4+i)},i=0,1,2,\ldots, n-4,\\
&0=\langle \varphi_3|M^{A^{\dag}}_{m}M^{A}_{m}\otimes I|\varphi_{i+n+2}\rangle=\langle 1+2|M^{A^{\dag}}_{m}M^{A}_{m}|(4+i)\rangle\\
&=a_{1(4+i)}+a_{2(4+i)}=a_{2(4+i)},i=0,1,2,\ldots, n-4,\\
&0=\langle \varphi_4|M^{A^{\dag}}_{m}M^{A}_{m}\otimes I|\varphi_5\rangle=\langle 2|M^{A^{\dag}}_{m}M^{A}_{m}|3+4\rangle=a_{23}+a_{24}=a_{23},\\
&0=\langle \varphi_2|M^{A^{\dag}}_{m}M^{A}_{m}\otimes I|\varphi_{i+n+2}\rangle=\langle 2+3|M^{A^{\dag}}_{m}M^{A}_{m}|(4+i)\rangle\\
&=a_{2(4+i)}+a_{3(4+i)}=a_{3(4+i)},i=0,1,2,\ldots, n-4,\\
&0=\langle \varphi_{i+n+2}|M^{A^{\dag}}_{m}M^{A}_{m}\otimes I|\varphi_{j+n+2}\rangle=a_{(4+i)(4+j)},i,j=0,1,2,\ldots, n-4,i\neq j.\\
\end{split}
\end{equation}

By the symmetry $\langle \xi_{i}|M^{A^{\dag}}_{m}M^{A}_{m}\otimes I|\xi_{j}\rangle=\langle \xi_{j}|M^{A^{\dag}}_{m}M^{A}_{m}\otimes I|\xi_{i}\rangle=0$, we immediately obtain that
$a_{ji}=a_{ij}=0 $ for all $1\leq i\neq j\leq n$.

Furthermore, we claim that $M^{A^{\dag}}_{m}M^{A}_{m}$ is a scalar matrix. It follows from $\langle \phi|M^{A^{\dag}}_{m}M^{A}_{m}\otimes I|\varphi_3\rangle=a_{11}+a_{12}-a_{21}-a_{22}=a_{11}-a_{22}=0$ 
that $a_{11}=a_{22}$. Similarly, for the quantum states $|\phi\rangle$ and $|\varphi_2\rangle$,
$|\phi\rangle$ and $|\varphi_5\rangle$, $|\phi\rangle$ and $|\varphi_{n+1}\rangle$, $|\phi\rangle$ and $|\varphi_{i+5}\rangle$ ($i=1,3,\ldots, 2\lfloor\frac{n-4}{2}\rfloor-1$), $|\phi\rangle$ and $|\varphi_{i+5}\rangle$ ($i=2,4,\ldots, 2\lceil\frac{n-4}{2}\rceil-2),$
we can deduce $a_{22}=a_{33}$, $a_{33}=a_{44}$, $a_{33}=a_{55}$ and $a_{(5+i)(5+i)}=a_{(6+i)(6+i)}$ ($i=0,1,2,\ldots, n-6).$

Therefore, $M^{A^{\dag}}_{m}M^{A}_{m}$ is proportional to the identity, which implies immediately that Alice cannot go first.

Next we consider the case that Bob goes first with a non-disturbing measurement $\{M^{B}_{m}\}$. We show that for any matrix $M^{B}_{m}$ preserving orthogonality of the states in $Eq.(1)$, $M^{B^{\dag}}_{m}M^{B}_{m}$ is proportional to the identity. Again write $M^{B^{\dag}}_{m}M^{B}_{m}=(b_{ij})$ where $i,j\in\{1,2,3,4\}$, which is clearly a symmetric matrix.
Similar to the case of $\{M^{A}_{m}\}$, we have that
$\langle \varphi_2|I\otimes M^{B^{\dag}}_{m}M^{B}_{m}|\varphi_3\rangle=\langle 3|M^{B^{\dag}}_{m}M^{B}_{m}|4\rangle=b_{34}=0$.
Likewise for the quantum states $|\varphi_{n+1}\rangle$ and $|\varphi_i\rangle, i=1,2,3,4,5$, we get

\begin{equation}
\begin{split}
&\langle \varphi_2|I\otimes M^{B^{\dag}}_{m}M^{B}_{m}|\varphi_5\rangle=\langle \varphi_5|I\otimes M^{B^{\dag}}_{m}M^{B}_{m}|\varphi_2\rangle=b_{13}=b_{31}=0,\\
&\langle \varphi_2|I\otimes M^{B^{\dag}}_{m}M^{B}_{m}|\varphi_{n+1}\rangle=\langle \varphi_{n+1}|I\otimes M^{B^{\dag}}_{m}M^{B}_{m}|\varphi_2\rangle=b_{23}=b_{32}=0,\\
&\langle \varphi_5|I\otimes M^{B^{\dag}}_{m}M^{B}_{m}|\varphi_{n+1}\rangle=\langle \varphi_{n+1}|I\otimes M^{B^{\dag}}_{m}M^{B}_{m}|\varphi_5\rangle=b_{12}=b_{21}=0,\\
&\langle \varphi_1|I\otimes M^{B^{\dag}}_{m}M^{B}_{m}|\varphi_3\rangle=\langle \varphi_3|I\otimes M^{B^{\dag}}_{m}M^{B}_{m}|\varphi_1\rangle\\
&=b_{24}+b_{34}=b_{42}+b_{43}=b_{42}=b_{24}=0,\\
&\langle \varphi_3|I\otimes M^{B^{\dag}}_{m}M^{B}_{m}|\varphi_4\rangle=\langle \varphi_4|I\otimes M^{B^{\dag}}_{m}M^{B}_{m}|\varphi_3\rangle\\
&=b_{14}+b_{24}=b_{41}+b_{42}=b_{41}=b_{14}=0.\\
\end{split}
\end{equation}

For each of $|\psi\rangle=|\varphi_4\rangle, |\varphi_1\rangle, |\varphi_{n+2}\rangle$, we have
$\langle \phi|I\otimes M^{B^{\dag}}_{m}M^{B}_{m}|\psi\rangle=0$, which implies that
$b_{11}=b_{22}=b_{33}=b_{44}$.
Hence $M^{B^{\dag}}_{m}M^{B}_{m}$ is proportional to the identity, which means that Bob cannot go first.\qed

We note that there are $8$ LOCC indistinguishable orthogonal product states in $\mathbb{C}^4\otimes\mathbb{C}^4$. Therefore the state $|\varphi_{n+1}\rangle=|3-5\rangle_A|2\rangle_B$ in (1) should be $|\varphi_{7}\rangle=|3+4\rangle_A|2\rangle_B$.

\bigskip
\noindent {\bf Theorem 2.}  In $\mathbb{C}^n\otimes\mathbb{C}^{2l}$ $(n \geq 2l > 4),$ there are $2(2l+n)-8$ 
 orthogonal product states that are LOCC indistinguishable and constructed as follows. 
\begin{equation}\label{e:2}
\begin{split}
&|\phi\rangle=|1-2+\cdots+(-1)^{n-1}n\rangle_A|1-2+\cdots-m\rangle_B,\\
&|\psi_i\rangle=|i+(i+1)\rangle_A|i\rangle_B,i=1,2,\ldots, l-1,\\
&|\psi_l\rangle=|(l-2)+(l-1)\rangle_A|l\rangle_B,\\
&|\psi_{i+l}\rangle=|i\rangle_A|l+(l+1)\rangle_B,i=1,2,\ldots, l-3,\\
&|\psi_{i+2l-3}\rangle=|(l+3+i)\rangle_A|l+(l+1)\rangle_B,i=1,2,\ldots, l-3,\\
&|\psi_{3l-5}\rangle=|l\rangle_A|l+(l+1)\rangle_B,\\
&|\psi_{3l-4}\rangle=|(l+2)\rangle_A|(l-1)+l\rangle_B,\\
&|\psi_{3l-3}\rangle=|(l+1)\rangle_A|(l-1)+l\rangle_B,\\
&|\psi_{i+3l-3}\rangle=|(2l+i)\rangle_A|l+(l+1)\rangle_B,i=1,2,\ldots, n-2l,\\
&|\varphi_i\rangle=|i\rangle_A|(2l-i)+(2l+1-i)\rangle_B,i=1,2,\ldots, l-1,\\
&|\varphi_{i+l-1}\rangle=|(2l+1-i)\rangle_A|i+(i+1)\rangle_B,i=1,2,\ldots, l-2,\\
&|\varphi_{2l-2}\rangle=|(l+1)+(l+2)\rangle_A|(l+1)\rangle_B,\\
&|\varphi_{2l-1}\rangle=|(l+3)\rangle_A|1+(l+1)\rangle_B,\\
&|\varphi_{i+2l-1}\rangle=|l+(l+1)\rangle_A|i\rangle_B,i=1,2,\ldots, l-2,\\
&|\varphi_{i+3l-3}\rangle=|l+(l+1)\rangle_A|(l+2+i)\rangle_B,i=1,2,\ldots, l-2,\\
& |\phi_i\rangle=\begin{cases}|(l+i)+(l+1+i)\rangle_A|(l+2)\rangle_B,& i=1,3,\ldots, 2\lfloor\frac{n-l}{2}\rfloor-1, \\
              |(l+i)+(l+1+i)\rangle_A|(l+3)\rangle_B,&i=2,4,\ldots, 2\lceil\frac{n-l}{2}\rceil-2.\end{cases}
\end{split}
\end{equation}
\noindent{ Proof:}
This conclusion can be proved similarly as Theorem 1. We first study the case that Alice goes first with a non-disturbing measurement $\{M^{A}_{m}\}$. Assume that the matrix $M^{A}_{m}$ preserves orthogonality of the states in \eqref{e:2}, i.e. if $M^{A}_{m}\otimes I|\xi_{i}\rangle$
are mutually orthogonal when $|\xi_{i}\rangle$ ranges from \eqref{e:2},  then $M^{A^{\dag}}_{m}M^{A}_{m}$ is the identity up to a constant. Write
$M^{A^{\dag}}_{m}M^{A}_{m}=(a_{ij})_{n\times n}$. For $1\leq i\neq j\leq n$ and
$i, j\neq l-2, l-1, l+3$, we have that
$a_{ij}=\langle \psi_{i+l}|M^{A^{\dag}}_{m}M^{A}_{m}\otimes I|\psi_{j+l}\rangle=0$. For the remaining indices, we have that

\begin{equation}
\begin{split}
&0=\langle \varphi_{2l-1}|M^{A^{\dag}}_{m}M^{A}_{m}\otimes I|\psi_{i+l}\rangle=a_{(l+3)i},\\
&0=\langle \varphi_{l-1}|M^{A^{\dag}}_{m}M^{A}_{m}\otimes I|\psi_{i+l}\rangle=a_{(l-1)i},\\
&0=\langle \psi_{l-1}|M^{A^{\dag}}_{m}M^{A}_{m}\otimes I|\psi_{3l-3}\rangle=a_{(l-1)(l+1)}+a_{l(l+1)}=a_{(l-1)(l+1)},\\
&0=\langle \varphi_{l-1}|M^{A^{\dag}}_{m}M^{A}_{m}\otimes I|\varphi_{2l-2}\rangle=a_{(l-1)(l+1)}+a_{(l-1)(l+2)}=a_{(l-1)(l+2)},\\
&0=\langle \varphi_{l-2}|M^{A^{\dag}}_{m}M^{A}_{m}\otimes I|\varphi_{l-1}\rangle=a_{(l-2)(l-1)},\\
&0=\langle \varphi_{l-1}|M^{A^{\dag}}_{m}M^{A}_{m}\otimes I|\varphi_{2l-1}\rangle=a_{(l-1)(l+3)},\\
&0=\langle \psi_{l}|M^{A^{\dag}}_{m}M^{A}_{m}\otimes I|\psi_{i+l}\rangle=a_{(l-1)i}+a_{(l-2)i}=a_{(l-2)i},\\
&0=\langle \psi_{l}|M^{A^{\dag}}_{m}M^{A}_{m}\otimes I|\varphi_{2l-1}\rangle=a_{(l-1)(l+3)}+a_{(l-2)(l+3)}=a_{(l-2)(l+3)}.
\end{split}
\end{equation}
Therefore $M^{A^{\dag}}_{m}M^{A}_{m}$ is a diagonal matrix.

For the quantum states $|\phi\rangle$, $|\varphi_{2l}\rangle$, $|\psi_i\rangle$ ($i=1,2,\ldots, l-1)$ and $|\phi_i\rangle$ ($i=1,2,\ldots, n-l-1$), we get that
$\langle i-(i+1)|M^{A^{\dag}}_{m}M^{A}_{m}|i+(i+1)\rangle=\langle i+(i+1)|M^{A^{\dag}}_{m}M^{A}_{m}|i-(i+1)\rangle=0$ ($i=1,2,\ldots, n-1$).
Thus $a_{ii}+a_{i(i+1)}-a_{(i+1)i}-a_{(i+1)(i+1)}=a_{ii}+a_{(i+1)i}-a_{i(i+1)}-a_{(i+1)(i+1)}=0$ ($i=1,2,\ldots, n-1$).
Since $a_{ij}=0 \ (i,j=1,2,\ldots, n, i\neq j)$, we have $a_{ii}=a_{(i+1)(i+1)}$ ($i=1,2,\ldots, n-1$).
Therefore $M^{A^{\dag}}_{m}M^{A}_{m}=a_{11}I$, which implies that Alice cannot go first.

Next we assume that Bob goes first with a non-disturbing measurement $\{M^{B}_{m}\}$.
If $M^{B}_{m}$ preserves the orthogonality of the states in \eqref{e:2}, we claim that
$M^{B^{\dag}}_{m}M^{B}_{m}=(b_{ij})$ is proportional to the identity.
We use the similar argument to see this. For $1\leq i\neq j\leq n$ and $i, j\neq l-1, l, l+1, l+2$, we have
that $b_{ij}=\langle \varphi_{i+2l-1}|I\otimes M^{B^{\dag}}_{m}M^{B}_{m}|\varphi_{j+2l-1}\rangle=0$. For the
other indices $i$ or $j=l-1, l, l+1, l+2$, the off-diagonal entries $b_{ij}=0$ are seen similarly as above.
For instance $\langle \psi_{l-l}|I\otimes M^{B^{\dag}}_{m}M^{B}_{m}|\varphi_{i+2l-1}\rangle=b_{(l-1)i}=0$.
The detailed computation for other cases go as follows.
\begin{equation}
\begin{split}
&0=\langle \psi_{l-l}|I\otimes M^{B^{\dag}}_{m}M^{B}_{m}|\psi_l\rangle=b_{(l-1)l},\\
&0=\langle \psi_{l-l}|I\otimes M^{B^{\dag}}_{m}M^{B}_{m}|\psi_{3l-5}\rangle=b_{(l-1)l}+b_{(l-1)(l+1)}=b_{(l-1)(l+1)},\\
&0=\langle \psi_{l-l}|I\otimes M^{B^{\dag}}_{m}M^{B}_{m}|\varphi_{l-1}\rangle=b_{(l-1)(l+1)}+b_{(l-1)(l+2)}=b_{(l-1)(l+2)},\\
&0=\langle \psi_{3l-3}|I\otimes M^{B^{\dag}}_{m}M^{B}_{m}|\varphi_{i+2l-1}\rangle=b_{(l-1)i}+b_{li}=b_{li},\\
&0=\langle \psi_{3l-3}|I\otimes M^{B^{\dag}}_{m}M^{B}_{m}|\varphi_{2l-2}\rangle=b_{(l-1)(l+1)}+b_{l(l+1)}=b_{l(l+1)},\\
&0=\langle \psi_{3l-3}|I\otimes M^{B^{\dag}}_{m}M^{B}_{m}|\phi_1\rangle=b_{(l-1)(l+2)}+b_{l(l+2)}=b_{l(l+2)},\\
&0=\langle \varphi_{2l-2}|I\otimes M^{B^{\dag}}_{m}M^{B}_{m}|\varphi_{i+2l-1}\rangle=b_{(l+1)i},\\
&0=\langle \varphi_{2l-2}|I\otimes M^{B^{\dag}}_{m}M^{B}_{m}|\phi_1\rangle=b_{(l+1)(l+2)},\\
&0=\langle \phi_1|I\otimes M^{B^{\dag}}_{m}M^{B}_{m}|\varphi_{i+2l-1}\rangle=b_{(l+2)i}.
\end{split}
\end{equation}

Therefore we have verified that $\langle \omega_{i}|I\otimes M^{B^{\dag}}_{m}M^{B}_{m}|\omega_{j}\rangle=\langle \omega_{j}|I\otimes M^{B^{\dag}}_{m}M^{B}_{m}|\omega_{i}\rangle=0$, where the states $|\omega_{i}\rangle$ belong to \eqref{e:2}. Therefore $M^{B^{\dag}}_{m}M^{B}_{m}$ is a diagonal matrix. 

To see it is a constant matrix, we consider
the quantum states $|\phi\rangle$, $|\varphi_{i}\rangle$ ($i=1,2,\ldots, 2l-3$), $|\psi_{3l-4}\rangle$ and $|\psi_{3l-5}\rangle$ which quickly imply that
$\langle i+(i+1)|M^{B^{\dag}}_{m}M^{B}_{m}|i-(i+1)\rangle=b_{ii}+b_{i(i+1)}-b_{(i+1)i}-b_{(i+1)(i+1)}=0$.
Therefore $b_{ii}=b_{(i+1)(i+1)}$ for $i=1,2,\ldots, m-1$. So $M^{B^{\dag}}_{m}M^{B}_{m}$ is a constant matrix.\qed

\bigskip
Case 2. In the case of $\mathbb{C}^n\otimes\mathbb{C}^m$($m=2k+1$), we find $d$ LOCC indistinguishable orthogonal product states,
where \begin{numcases}{d=}
 2(n-m)+9,&for $n \geq m = 5 $\\
  2(n+m)-7,&for $n \geq m > 5 $
 \end{numcases}
which are listed as follows.

 \begin{equation}\label{e:3}
\begin{split}
&|\phi\rangle=|1-2+\cdots+(-1)^{n-1}n\rangle_A|1-2+\cdots+m\rangle_B,\\
&|\phi_1\rangle=|(k-2)+(k-1)\rangle_A|k\rangle_B,\\
&|\phi_2\rangle=|(k+3)+(k+4)\rangle_A|(k+2)\rangle_B,\\
&|\phi_3\rangle=|k\rangle_A|(k+3)+(k+4)\rangle_B,\\
&|\phi_4\rangle=|(k+2)\rangle_A|(k-2)+(k-1)\rangle_B,\\
&|\psi_i\rangle=\begin{cases} |i+(i+1)\rangle_A|i\rangle_B,& i=1,2,\ldots, k,\\
         |i+(i+1)\rangle_A|(i+1)\rangle_B,& i=k+1,k+2,\ldots, 2k \end{cases},\\
&|\psi_{i+2k}\rangle=|i\rangle_A|(k+1)+(k+2)\rangle_B,i=1,2,\ldots, k-2,\\
&|\psi_{i+2k-5}\rangle=|i\rangle_A|k+(k+1)\rangle_B,i=k+4,k+5,\ldots, n,\\
&|\varphi_i\rangle=\begin{cases} |i\rangle_A|(2k+1-i)+(2k+2-i)\rangle_B,&i=1,2,\ldots, k,\\
                  |(i+1)\rangle_A|(2k+1-i)+(2k+2-i)\rangle_B,&i=k+1,k+2,\ldots, 2k \end{cases} \\
&|\varphi_{i+2k}\rangle=|k+(k+1)\rangle_A|i\rangle_B,i=1,2,\ldots, k-2,\\
&|\varphi_{i+2k-5}\rangle=|(k+1)+(k+2)\rangle_A|i\rangle_B,i=k+4,k+5,\ldots, 2k+1,\\
&|\phi_{i+5}\rangle=\begin{cases} |(2k+i)+(2k+1+i)\rangle_A|2k\rangle_B,& i=1,3,\ldots, 2\lceil\frac{n-m}{2}\rceil-1, \\
        |(2k+2+i)+(2k+3+i)\rangle_A|(2k+1)\rangle_B,&i=0,2,\ldots, 2\lfloor\frac{n-m}{2}\rfloor-2.\end{cases}
\end{split}
\end{equation}

\noindent{\bf Theorem 3.}  In $\mathbb{C}^n\otimes\mathbb{C}^{2k+1}$ $(n \geq 2k+1 \geq 5 )$, there are $d$ LOCC indistinguishable
orthogonal product given in \eqref{e:3}.

\bigskip
\noindent{ Proof:}
This result is proved by the similar method. One shows that
the orthogonal product states in \eqref{e:3} are indistinguishable by LOCC no matter who goes first.
Using the obvious symmetry when $m=n$,
we only need to check these states cannot be distinguished by LOCC when Alice goes first.
As before we assume Alice goes first with a non-disturbing measurement $\{M^{A}_{m}\}$. Write $M^{A^{\dag}}_{m}M^{A}_{m}=(a_{ij})$ for $1\leq i, j\leq n$.  Using the property that
$M^{A}_m$ preserves the orthogonality of the states in \eqref{e:3}, we can see that $(a_{ij})$ is a diagonal matrix.

As this is quite similar to the last theorem, we list below the results of applying
$M_m^A$.
\begin{itemize}
\item $|\varphi_{k}\rangle$, $|\varphi_{k+1}\rangle$ and $|\psi_{i+2k}\rangle$  ($i=1,2,\ldots, n-5$)
imply that $a_{ij}=0$ for $i,j=1,2,\ldots, k-2,k,k+2,k+4,\ldots, n, i\neq j$.
\item $|\psi_{k}\rangle$, $|\psi_{k+1}\rangle$, $|\varphi_{i}\rangle$ ($i=k-1,k,k+1,k+2$) and $|\psi_{i+2k}\rangle$ ($i=1,2,\ldots, n-5$)
imply that $a_{(k+1)i}=a_{i(k+1)}=0 (i=1,2,\ldots, k,k+2,\ldots, n)$.
\item $|\phi_{1}\rangle$, $|\psi_{k+1}\rangle$, $|\varphi_{i}\rangle$ ($i=k-1,k,k+1,k+2$) and $|\psi_{i+2k}\rangle$ ($i=1,2,\ldots, n-5$)
imply that $a_{(k-1)i}=a_{i(k-1)}=0 (i=1,2,\ldots, k-2,k,\ldots, n)$.
\item $|\phi_{2}\rangle$, $|\psi_{k}\rangle$, $|\varphi_{i}\rangle$ ($i=k-1,k,k+1,k+2$) and $|\psi_{i+2k}\rangle$ ($i=1,2,\ldots, n-5$)
imply that $a_{(k+3)i}=a_{i(k+3)}=0 (i=1,2,\ldots, k+2,k+4,\ldots, n)$.
\end{itemize}

Therefore $(a_{ij})$ is diagonal. 
Using the quantum states $|\phi\rangle$, $|\psi_{i}\rangle$ ($i=1,2,\ldots, 2k$) and $|\phi_{i+5}\rangle (i=0,1,2,\ldots, n-m-1),$
it is not hard to see that $\langle i+(i+1)|M^{A^{\dag}}_{m}M^{A}_{m}|i-(i+1)\rangle=a_{ii}-a_{(i+1)(i+1)}=0 \ (i=1,2,\ldots, n-1).$
Therefore $M^{A^{\dag}}_{m}M^{A}_{m}$ is a constant matrix, so Alice cannot go first.\qed

\section{Distinguishing orthogonal product states with the least entanglement resource}

In this section we show that the indistinguishable orthogonal product states given in the last section can be distinguished by LOCC using a $\mathbb{C}^2\otimes\mathbb{C}^2$ maximally entangled state as a resource.

\bigskip
\noindent {\bf Theorem 4.} In $\mathbb{C}^n\otimes\mathbb{C}^m$ $(n > m = 4)$, it is sufficient to perfectly distinguish the $2n-1$ orthogonal product states in (1) by LOCC, using a $\mathbb{C}^2\otimes\mathbb{C}^2$ MES.\\

\noindent{ Proof:}
To distinguish these states by LOCC, we add two
auxiliary $2$ dimensional systems $\mathcal{H}^{a}$ and $\mathcal{H}^{b}$ and consider the maximally entangled state $|\psi\rangle_{ab}=\frac{1}{\sqrt{2}}\sum\limits_{i=1}^{2}|ii\rangle_{ab}$.
 Alice and Bob have access to the subsystems $aA$ and $bB$ respectively.

Alice starts to make a two-outcome measurement

$$\emph{A}_1=|1\rangle_{a}\langle 1|\otimes|1\rangle_{A}\langle 1|+|2\rangle_{a}\langle 2|\otimes(|2\rangle_{A}\langle 2|+|3\rangle_{A}\langle 3|+\cdots+|n\rangle_{A}\langle n|),$$
$$\emph{A}_2=|2\rangle_{a}\langle 2|\otimes|1\rangle_{A}\langle 1|+|1\rangle_{a}\langle 1|\otimes(|2\rangle_{A}\langle 2|+|3\rangle_{A}\langle 3|+\cdots+|n\rangle_{A}\langle n|).$$

By operating $\emph{A}_1(|\omega_i\rangle_{AB}\otimes|\psi\rangle_{ab})$, where $|\omega_i\rangle_{AB}$ belongs to \eqref{e:1}, we get a new set of
states as follows.

\begin{equation}
\begin{split}
&|\phi'\rangle=(|1\rangle_{A}|11\rangle_{ab}+|-2+\cdots+(-1)^{n-1}n\rangle_A|22\rangle_{ab})|1-2+3-4\rangle_B,\\
&|\varphi'_1\rangle=|\varphi_1\rangle|11\rangle_{ab},\\
&|\varphi'_2\rangle=|\varphi_2\rangle|22\rangle_{ab}, \\
&|\varphi'_3\rangle=(|1\rangle_A|11\rangle_{ab}+|2\rangle_A|22\rangle_{ab})|4\rangle_B, \\
&|\varphi'_{i}\rangle=|\varphi_{i}\rangle|22\rangle_{ab},i=4,5\ldots, 2n-2.\\
\end{split}
\end{equation}

We claim that the states can be distinguished in $\mathcal{H}^{aA}\otimes\mathcal{H}^{bB}$. In fact, Bob makes a $3$-outcome projective measurement. His first outcome $B_1=|1\rangle_{b}\langle 1|\otimes|2+3\rangle_{B}\langle 2+3|$ leaves $|\varphi'_1\rangle$ invariant, so $|\varphi'_1\rangle$ is successfully identified.

Bob's second outcome $B_2=|2\rangle_{b}\langle 2|\otimes|1+2\rangle_{B}\langle 1+2|$ leaves $|\varphi'_i\rangle$ ($i=4,5,\ldots, n+1$)
invariant. To further identify the states, Alice uses the projector $A_{21}=|2\rangle_{a}\langle 2|\otimes|2\rangle_{A}\langle 2|$, which keeps $|\varphi'_4\rangle$ invariant. That is to say, $|\varphi'_4\rangle$ has been successfully identified. Then Alice uses the projector $A_{22}=|2\rangle_{a}\langle 2|\otimes(|3\rangle_{A}\langle 3|+|4\rangle_{A}\langle 4|+\cdots+|n\rangle_{A}\langle n|)$, which leaves $|\varphi'_i\rangle$ ($i=5,6,\ldots, n+1$) invariant. Alice and Bob can further distinguish them by LOCC accordingly.

Bob's last outcome $B_{3}=|1\rangle_{b}\langle 1|\otimes|4\rangle_{B}\langle 4|+|2\rangle_{b}\langle 2|\otimes(|3\rangle_{B}\langle 3|+|4\rangle_{B}\langle 4|)$ keeps $|\varphi'_{2}\rangle$, $|\varphi'_{3}\rangle$ and $|\varphi'_{i}\rangle$ ($i=n+2,n+3,\ldots, 2n-2$) invariant and transforms $|\phi'\rangle$ to $|1\rangle_A|11\rangle_{ab}|4\rangle_{B}+|-2+3-4+\cdots+(-1)^{n-1}n\rangle_A|22\rangle_{ab}|3-4\rangle_{B}$.
Then Alice uses the projector $A_{31}=|1\rangle_{a}\langle 1|\otimes|1\rangle_{A}\langle 1|+|2\rangle_{a}\langle 2|\otimes(|2\rangle_{A}\langle 2|+|3\rangle_{A}\langle 3|)$, which leaves $|\varphi'_{2}\rangle$, $|\varphi'_{3}\rangle$ invariant and transforms $|\phi'\rangle$ to $|1\rangle_A|11\rangle_{ab}|4\rangle_{B}+|-2+3\rangle_A|22\rangle_{ab}|3-4\rangle_{B}$. Hence these states can be easily distinguished by Alice and Bob using LOCC. When Alice uses the projectors $A_{3i}=2\rangle_{a}\langle 2|\otimes|2+i\rangle_{A}\langle 2+i|$ for $i=2,3,\ldots, n-2$, the other states $|\varphi'_{i}\rangle$ ($i=n+2,n+3,\ldots, 2n-2$) can be distinguished easily.\qed

\bigskip
\noindent{\bf Theorem 5.} In $\mathbb{C}^n\otimes\mathbb{C}^{2l}$ $(n \geq 2l > 4)$, it is sufficient to perfectly distinguish the $2(2l+n)-8$ orthogonal product states in \eqref{e:2} by LOCC via a $\mathbb{C}^2\otimes\mathbb{C}^2$ MES.

\bigskip
\noindent{ Proof:}
As the proof of Theorem 4, Alice and Bob first share a maximally entangled state $|\psi\rangle_{ab}=\frac{1}{\sqrt{2}}\sum\limits_{i=1}^{2}|ii\rangle_{ab}$. Alice prepares the two-outcome measurement

$$\emph{A}_1=|1\rangle_{a}\langle 1|\otimes\sum\limits_{i=1}^{l-1}|i\rangle_{A}\langle i|+|2\rangle_{a}\langle 2|\otimes\sum\limits_{i=l}^{m}|i\rangle_{A}\langle i|,$$
$$\emph{A}_2=|2\rangle_{a}\langle 2|\otimes\sum\limits_{i=1}^{l-1}|i\rangle_{A}\langle i|+|1\rangle_{a}\langle 1|\otimes\sum\limits_{i=l}^{m}|i\rangle_{A}\langle i|.$$

Upon applying $\emph{A}_1(|\omega_i\rangle_{AB}\otimes|\psi\rangle_{ab})$, where $|\omega_i\rangle_{AB}$ are the states in \eqref{e:2}, a new set of
states is obtained as follows.

\begin{equation}
\begin{split}
&|\phi'\rangle=(|\sum\limits_{i=1}^{l-1}(-1)^{i-1}i\rangle_{A}|11\rangle_{ab}+|\sum\limits_{i=l}^{n}(-1)^{i-1}i\rangle_{A}|22\rangle_{ab})|1-2+\cdots+(-1)^{m-1}m\rangle_B\\
&|\psi'_i\rangle=|\psi_i\rangle|11\rangle_{ab},i=1,2,\ldots, l-2,\\
&|\psi'_{l-1}\rangle=(|(l-1)\rangle_A|11\rangle_{ab}+|l\rangle_A|22\rangle_{ab})|(l-1)\rangle_B,\\
&|\psi'_{i+l}\rangle=|\psi_{i+l}\rangle|11\rangle_{ab},i=0,1,2,\ldots, l-3,\\
&|\psi'_i\rangle=|\psi_i\rangle|22\rangle_{ab},i=2l-2,2l-1\ldots, n+l-3,\\
&|\varphi'_i\rangle=|\varphi_i\rangle|11\rangle_{ab},i=1,2,\ldots, l-1,\\
&|\varphi'_i\rangle=|\varphi_i\rangle|22\rangle_{ab},i=l,l+1\ldots, 4l-5,\\
&|\phi'_i\rangle=|\phi_i\rangle|22\rangle_{ab},i=1,2,\ldots, n-l-1. \\
\end{split}
\end{equation}

Now we need to show that these states can be distinguished in $\mathcal{H}^{aA}\otimes\mathcal{H}^{bB}$. Bob makes an $(m-1)$-outcome projective measurement. The first outcome
$B_1=|1\rangle_{b}\langle 1|\otimes|1\rangle_{B}\langle 1|$ leaves $|\psi'_{1}\rangle$ invariant, while transforms $|\phi'\rangle$ to $|\sum\limits_{j=1}^{l-1}(-1)^{j-1}j\rangle_{A}|11\rangle_{ab}|1\rangle_{B}$. They can be easily distinguished by Alice by projection onto $|l\pm 2\rangle_{A}$.
In the same way, the outcome $B_i=|1\rangle_{b}\langle 1|\otimes|i\rangle_{B}\langle i|$ ($i=2,3,\ldots, l-2)$ keeps $|\psi'_{i}\rangle$ invariant and transforms $|\phi'\rangle$ to $|\sum\limits_{j=1}^{l-1}(-1)^{j-1}j\rangle_{A}|11\rangle_{ab}|i\rangle_{B}$ ($i=2,3,\ldots, l-2$). They are also
easily distinguished by Alice via projecting onto $|i\pm (i+1)\rangle_{A}$.

For the outcome $B_{l-1}=|1\rangle_{b}\langle 1|\otimes(|l\rangle_{B}\langle l|+|(l+1)\rangle_{B}\langle (l+1)|+\cdots+|m\rangle_{B}\langle m|$, it leaves $|\psi'_i\rangle$ ($i=l,l+1,\ldots, 2l-3$), $|\varphi'_{i}\rangle$ ($i=1,2,\ldots, l-1$) invariant and transforms $|\phi'\rangle$ to $|\sum\limits_{j=1}^{l-1}(-1)^{j-1}j\rangle_{A}|11\rangle_{ab}|\sum\limits_{j=l}^{m}(-1)^{j-1}j\rangle_B$. Then Alice uses a rank one
projector $A_{(l-1)1}=|1\rangle_{a}\langle 1|\otimes|1\rangle_{A}\langle 1|$, which leaves $|\psi'_{l+1}\rangle,$ $|\varphi'_{1}\rangle$ invariant, while transforms $|\phi'\rangle$ to $|1\rangle_{A}|11\rangle_{ab}|\sum\limits_{j=l}^{m}(-1)^{j-1}j\rangle_B$. These states can be easily identified by Bob. Similarly, when Alice uses the projector $A_{(l-1)i}=|1\rangle_{a}\langle 1|\otimes|i\rangle_{A}\langle i|$ ($i=2,3,\ldots, l-3$), it leaves $|\psi'_{l+i}\rangle,$ $|\varphi'_{i}\rangle$ intact, and transforms $|\phi'\rangle$ to $|i\rangle_{A}|11\rangle_{ab}|\sum\limits_{j=l}^{m}(-1)^{j-1}j\rangle_B$  ($i=2,3,\ldots, l-3$). These
states can also be distinguished by Bob. When Alice uses the projector $A_{(l-1)(l-2)}=|1\rangle_{a}\langle 1|\otimes|(l-2)+(l-1)\rangle_{A}\langle (l-2)+(l-1)|$, it leaves $|\psi'_{l}\rangle,$ $|\varphi'_{l-2}\rangle$ and $|\varphi'_{l-1}\rangle$ invariant. They are further
distinguished by Alice and Bob using LOCC.

For the outcome $B_{l}=|1\rangle_{b}\langle 1|\otimes|(l-1)\rangle_{B}\langle (l-1)|+|2\rangle_{b}\langle 2|\otimes(|1\rangle_{B}\langle 1|+|2\rangle_{B}\langle 2|+\cdots+|(l+1)\rangle_{B}\langle (l+1)|$, it leaves $|\psi'_{l-1}\rangle,$ $|\psi'_{i}\rangle$ $(i=2l-2,2l-1,\ldots, n+l-3)$, and $|\varphi'_{i}\rangle$ $(i=l,l+1,\ldots, 3l-3)$ invariant, while transforms $|\phi'\rangle$ to $|\sum\limits_{i=1}^{l-1}(-1)^{i-1}i\rangle_{A}|11\rangle_{ab}|(-1)^{l-1}(l-1)\rangle_B+|\sum\limits_{i=l}^{n}(-1)^{i-1}i\rangle_{A}|22\rangle_{ab}|1-2+\cdots+(-1)^{l+1}(l+1)\rangle_B$.
Then Alice uses the projector $A_{l1}=|1\rangle_{a}\langle 1|\otimes|(l-1)\rangle_{A}\langle (l-1)|+|2\rangle_{a}\langle 2|\otimes(|l\rangle_{A}\langle l|+|(l+1)\rangle_{A}\langle (l+1)|+|(l+2)\rangle_{A}\langle (l+2)|)$, which leaves $|\psi'_{l-1}\rangle,$ $|\psi'_{3l-5}\rangle,$ $|\psi'_{3l-4}\rangle$, $|\psi'_{3l-3}\rangle,$ $|\varphi'_{2l-2}\rangle,$ $|\varphi'_{2l-1}\rangle$ and $|\varphi'_{i+2l-1}\rangle$ ($i=1,2,\ldots, l-2$) invariant and transforms $|\phi'\rangle$ to $|(-1)^{l-1}(l-1)\rangle_{A}|11\rangle_{ab}|(-1)^{l-1}(l-1)\rangle_B+|l-(l+1)+(l+2)\rangle_{A}|22\rangle_{ab}|1-2+\cdots+(-1)^{l+1}(l+1)\rangle_B$.
Then Alice and Bob can distinguish these states by LOCC. For other states, Alice using the projectors $A_{li}=|2\rangle_{a}\langle 2|\otimes|(l+1+i)\rangle_{A}\langle (l+1+i)|$ ($i=2,3,\ldots, n-l-1$), Bob can easily distinguish them accordingly.

Similarly, for the remaining states $|\varphi'_{i+3l-3}\rangle$ ($i=1,2,\ldots, l-2$) and $|\phi'_{i}\rangle$ ($i=1,2,\ldots, n-l-1$), Bob uses the projectors $B_{l+i}=|2\rangle_{b}\langle 2|\otimes|(l+1+i)\rangle_{B}\langle (l+1+i)|$ ($i=1,2,\ldots, m-l-1$), then Alice can easily distinguish them.\qed

\bigskip
\noindent {\bf Theorem 6.} In $\mathbb{C}^n\otimes\mathbb{C}^{2k+1}$ $(n \geq 2k+1 \geq 5 ),$ it is sufficient to perfectly distinguish the $d$ orthogonal product states in $Eq.(9)$ by LOCC using only a $\mathbb{C}^2\otimes\mathbb{C}^2$ MES, where
\begin{numcases}{d=}
 2(n-m)+9,&for $n \geq m = 5 $\\
  2(n+m)-7,&for $n \geq m > 5 $
 \end{numcases}

\noindent{ Proof:}
 To distinguish these states, Alice and Bob first share a maximally entangle state $|\psi\rangle_{ab}$, where $|\psi\rangle_{ab}=\frac{1}{\sqrt{2}}\sum\limits_{i=1}^{2}|ii\rangle_{ab}$.
Then, Bob makes a two-outcome measurement

$$\emph{B}_1=|1\rangle_{b}\langle 1|\otimes\sum\limits_{i=1}^{k+1}|i\rangle_{B}\langle i|+|2\rangle_{b}\langle 2|\otimes\sum\limits_{i=k+2}^{m}|i\rangle_{B}\langle i|,$$
$$\emph{B}_2=|2\rangle_{b}\langle 2|\otimes\sum\limits_{i=1}^{k+1}|i\rangle_{B}\langle i|+|1\rangle_{b}\langle 1|\otimes\sum\limits_{i=k+2}^{m}|i\rangle_{B}\langle i|.$$

By operating $\emph{B}_1(|\omega_i\rangle_{AB}\otimes|\psi\rangle_{ab})$, where $|\omega_i\rangle_{AB}$ belongs to \eqref{e:3}, we get a new set of
states as follows.

 \begin{equation}
\begin{split}
&|\phi'\rangle=|1-2+\cdots+(-1)^{n-1}n\rangle_A(|\sum\limits_{i=1}^{k+1}(-1)^{i-1}i\rangle_{B}|11\rangle_{ab}+|\sum\limits_{i=k+2}^{m}(-1)^{i-1}i\rangle_{B}|22\rangle_{ab})\\
&|\phi'_{1,4}\rangle=|\phi_{1,4}\rangle|11\rangle_{ab},~~~~~~~~~~|\phi'_{2,3}\rangle=|\phi_{2,3}\rangle|22\rangle_{ab},\\
&|\psi'_i\rangle=|\psi_i\rangle|11\rangle_{ab},i=1,2,\ldots, k,\\
&|\psi'_i\rangle=|\psi_i\rangle|22\rangle_{ab},i=k+1,k+2,\ldots, 2k,\\
&|\psi'_{i+2k}\rangle=|i\rangle_A(|(k+1)\rangle_B|11\rangle_{ab}+|(k+2)\rangle_B|22\rangle_{ab}),i=1,2,\ldots, k-2,\\
&|\psi'_{i+2k-5}\rangle=|\psi_{i+2k-5}\rangle|11\rangle_{ab},i=k+4,k+5,\ldots, n,\\
&|\varphi'_i\rangle=|\varphi_i\rangle|22\rangle_{ab},i=1,2,\ldots, k-1,\\
&|\varphi'_k\rangle=|k\rangle_A(|(k+1)\rangle_B|11\rangle_{ab}+|(k+2)\rangle_B|22\rangle_{ab}),\\
&|\varphi'_i\rangle=|\varphi_i\rangle|11\rangle_{ab},i=k+1,k+2,\ldots, 2k,\\
&|\varphi'_{i+2k}\rangle=|\varphi_{i+2k}\rangle|11\rangle_{ab},i=1,2,\ldots, k-2,\\
&|\varphi'_{i+2k-5}\rangle=|\varphi_{i+2k-5}\rangle|22\rangle_{ab},i=k+4,k+5,\ldots, m,\\
&|\phi'_{i+5}\rangle=|\phi_{i+5}\rangle|22\rangle_{ab},i=0,1,2,\ldots, n-m-1. \\
\end{split}
\end{equation}

Now we need to show that the states can be distinguished in $\mathcal{H}^{aA}\otimes\mathcal{H}^{bB}$. Alice makes an $(n-k+2)$-outcome projective measurement, in which the first outcome
$A_1=|1\rangle_{a}\langle 1|\otimes|(k+3)\rangle_{A}\langle (k+3)|$ leaves $|\varphi'_{k+2}\rangle$ intact and transforms $|\phi'\rangle$ to $|(k+3)\rangle_A|\sum\limits_{i=1}^{k}(-1)^{i-1}i\rangle_{B}|11\rangle_{ab}$. Therefore Bob can easily distinguish the results by applying them to $|(k-1)\pm k\rangle_{B}$.

Alice's second outcome $A_2=|2\rangle_{a}\langle 2|\otimes|(k-1)\rangle_{A}\langle (k-1)|$ keeps $|\varphi'_{k-1}\rangle$ invariant, while transforms
$|\phi'\rangle$ to $|(k-1)\rangle_A|\sum\limits_{i=k+2}^{m}(-1)^{i-1}i\rangle_{B}|22\rangle_{ab}$, which are easily distinguished by Bob via projections on $|(k+2)\pm (k+3)\rangle_{B}$.

The third outcome $A_3=|1\rangle_{a}\langle 1|\otimes|(k+2)\rangle_{A}\langle (k+2)|$ leaves $|\phi'_{4}\rangle$ and $|\varphi'_{k+1}\rangle$ invariant and maps $|\phi'\rangle$ to $|(k+2)\rangle_A|\sum\limits_{i=1}^{k+1}(-1)^{i-1}i\rangle_{B}|11\rangle_{ab}$. Then Bob uses the
projectors $B_{31,32}=|1\rangle_{b}\langle 1|\otimes|(k-2)\pm(k-1)\rangle_{B}\langle (k-2)\pm(k-1)|$ and $B_{33,34}=|1\rangle_{b}\langle 1|\otimes|k\pm(k+1)\rangle_{B}\langle k\pm(k+1)|$ to identify the states $|\phi'_{4}\rangle$, $|\varphi'_{k+1}\rangle$ and $|\phi'\rangle$.
In the same way, the outcomes $A_{i+3}=|1\rangle_{a}\langle 1|\otimes|(k+3+i)\rangle_{A}\langle (k+3+i)|$ ($i=1,2,\ldots, n-k-3$)
can distinguish the states $|\varphi'_{i}\rangle$ ($i=k+3,k+4,\ldots, 2k$) and $|\psi'_{i+2k-5}\rangle (i=k+4,k+5,\ldots, n)$.

Using the same method as the proof of Theorem 5, Alice's outcome $A_{n-k+1}=|2\rangle_{a}\langle 2|\otimes(|(k+1)\rangle_{A}\langle (k+1)|+|(k+2)\rangle_{A}\langle (k+2)|+\cdots+|n\rangle_{A}\langle n|)$ help to distinguish the states $|\phi'_{2}\rangle$, $|\psi'_{i}\rangle$ ($i=k+1,k+2,\ldots, 2k$), $|\varphi'_{i+2k-5}\rangle$ ($i=k+4,k+5,\ldots, 2k+1$) and $|\phi'_{i+5}\rangle$ ($i=0,1,2,\ldots, n-m-2$).

Alice's last outcome is $A_{n-k+2}=|1\rangle_{a}\langle 1|\otimes(|1\rangle_{A}\langle 1|+|2\rangle_{A}\langle 2|+\cdots+|(k+1)\rangle_{A}\langle (k+1)|)+|2\rangle_{a}\langle 2|\otimes(|1\rangle_{A}\langle 1|+|2\rangle_{A}\langle 2|+\cdots+|(k-2)\rangle_{A}\langle (k-2)|+|k\rangle_{A}\langle k|)$. When Bob uses the projectors $B_{(n-k+2)i}=|1\rangle_{b}\langle 1|\otimes|i\rangle_{B}\langle i|$ for $i=1,2,\ldots, k$, Alice can easily distinguish the states $|\phi'_{1}\rangle$, $|\psi'_{i}\rangle$ ($i=1,2,\ldots, k$) and $|\varphi'_{i+2k}\rangle$ ($i=1,2,\ldots, k-2$).
When Bob uses the projector $B_{(n-k+2)(k+1)}=|1\rangle_{b}\langle 1|\otimes|(k+1)\rangle_{B}\langle (k+1)|+|2\rangle_{b}\langle 2|\otimes|(k+2)\rangle_{B}\langle (k+2)|,$ the states $|\varphi'_{k}\rangle$ and $|\psi'_{i+2k}\rangle$ ($i=1,2,\ldots, k-2$) can be distinguished by Alice.
When Bob uses the projector $B_{(n-k+2)(k+2)}=|2\rangle_{b}\langle 2|\otimes(|(k+2)\rangle_{B}\langle (k+2)|+|(k+3)\rangle_{B}\langle (k+3)|+\cdots+|(2k+1)\rangle_{B}\langle (2k+1)|),$ the states $|\phi'_{3}\rangle$ and $|\varphi'_{i}\rangle$ ($i=1,2,\ldots, k-2$) are
identified.\qed

\section{Discussions and Conclusion}

In this paper, we have studied how to distinguish orthogonal product states by LOCC with the least entanglement resource.
We have constructed a new set of locally indistinguishable orthogonal product states in $\mathbb{C}^n\otimes\mathbb{C}^m(n \geq m \geq 4)$
and proved their local distinguishability by using a $\mathbb{C}^2\otimes\mathbb{C}^2$ maximally entangled state. Our result helps to
understand further the phenomenon of nonlocality without entanglement.

It is clear that not any set of bipartite quantum states can be locally distinguished, if Alice and Bob share only $1$-qubit entanglement. It is interesting to know what entanglement resources are sufficient and optimal for a given set of indistinguishable states.

$Acknowledgments:$ This work is supported by the National Natural Science Foundation of China (Grant No. 11571119, 11501153, 11701504, 11531004), Simons Foundation (Grant no. 523868) and
the Young Innovative Talent Project of Department of Education of Guangdong Province (Grant No. 2016KQNCX180).

\end {document}